# Satellite Based IoT Networks for Emerging Applications

Sudhir K. Routray
Department of Electrical and Computer Engineering
College of Electircal and Mechanical Engineering
Addis Ababa Science and Technology University
Addis Ababa, Ethiopia
Email: sudhir.routray@aastu.edu.et

Habib Mohammed Hussein
Department of Electrical and Computer Engineering
College of Electircal and Mechanical Engineering
Addis Ababa Science and Technology University
Addis Ababa, Ethiopia
Email: habib.mohammed@aastu.edu.et

*Abstract*—Internet of things (IoT) has been proven to be a ubiquitous technology for several existing and new applications. It can provide both accuracy and sustainability in the emerging services and applications. IoT has several advantages and it can provide different levels of coverage in different geographical locations and contexts. There are several applications in which the wide coverage, low power and reliability of the communication have very high priorities. Conglomeration of IoT and satellite networks can make it very attractive for these applications. In this article, we present the general features, motivation and deployment issues of satellite based IoT networks. We present the applications where these satellite based IoT networks can play important roles. We also present the hybridization of narrowband IoT and satellite networks which can provide a long term sustainable network solution for several applications.

*Keywords—Internet of things; satellite based IoT; applications of satellite based IoT; narrowband IoT; Green IoT*

## I. INTRODUCTION

Internet of things (IoT) is pervasive extension of the Internet in different domains. In fact, the dimension of the Internet can be expanded in many ways using IoTs. Both human and machine can communicate using IoT based frameworks. Overall, data transfer between both living and nonliving entities can be made possible through IoT. It uses several sensors based on the application requirements for connecting them together. Now IoT has several dimensions for numerous applications and it facilitates massive machine type communications. IoT can reach to the territories where there is no Internet available. Internet is normally available where there are human settlements. However, there are many scenarios where we need the information without the presence of the Internet. For instance, measurements of under surface soil humidity and fertilizer intensity can be brought in to the Internet through the IoT. This can be done by deploying appropriate sensors under the surface. Similarly, there are applications in which micro monitoring is required in the unmanned territories. In those cases, the supervision of the IoT can be done through some remote communication nodes. Satellites and high altitude based communication devices are instrumental in these applications.

There are several works have been done on the recent advances of IoT. In fact, IoT has already been deployed in several fields of applications and the emerging applications are being tested for the future deployments. Applications of IoTs for the industrial purposes and their impact have been surveyed in [1]. Using IoT several industrial functions can be automated. In [2], a comprehensive survey on IoTs and their enabling technologies have been presented. It clearly mentions the needs and outcomes of the IoT based systems. In [3], current figures and statistics of modern communication world are presented which shows the device densities in use in the recent years. Using both the mobile and fixed infrastructure IoT can be realized for sensing and ts associated functions. A new method for forward link based satellite IoT (SIoT) system has been proposed in [4]. In this system the effectiveness of the overall SIoT system has been analyzed taking the physical parameters in to account. A low earth orbit (LEO) constellation has been proposed for IoT related applications in [5]. In this work the main requirements and compatibility of both satellites and the IoTs have been studied for the SIoT framework. An architecture for LEO based SIoT system has been proposed. Spectrum allocations, constellation structure of the LEO satellites, compatibility between the heterogeneous parts in the SIoT network, and the protocols for the proposed SIoT have been studied in [5]. The number of sensors used in IoT networks is too high. Therefore, energy efficiency in IoTs is a main concern for the recent applications. In [6] and [7] green IoTs are considered for the practical applications and deployment. These IoTs are good for energy limited applications and the long term sustainability. Narrowband IoT (NBIoT) is one of the energy efficient forms of IoTs [6]. It is very much effective in the resource limited applications. NBIoT has several attractive features which makes it suitable for green applications including SIoT [6] - [8]. There are several emerging applications of SIoT. The main SIoT applications include the mission critical applications where the network coverage has to be there all the time, location determination, navigation monitoring, and location based services [9] – [16]. Every year new applications are getting added to the pool of applications of SIoT. Security of IoT is a challenging aspect due to its vulnerabilities from several grounds. In [17], a robust security for IoT has been proposed for diverse scenarios. In [18], main motivations for SIoT have been presented. It shows that satellite based IoTs provide several advantages over the cellular IoTs in terms of resource saving and coverage.

In this article, we present the emergence of SIoT networks as a new dimension to both the satellites and IoTs. We show that this network has specific importance for the future. We present the sustainability aspects of SIoT and its advantages in the long term.

The reminder of this article is organized in four different sections. In section II, we provide the motivation and essence of SIoT networks. Later in the section, we provide the dual mode satellite cellular IoT networks. In section III, we present the applications of SIoT. In section IV, we discuss the sustainability of SIoT and the emerging solutions for it. We suggest that NBIoT is a suitable solution for SIoT, In section V, we conclude the article with the main points of the paper.

## II. Satellite Based IoT Networks

Normally, satellites and IoT do not seem to be natural partners in the communication world. However, due to the compulsions of some specific situations they meet each other in several instances. Therefore there are several reasons the IoT should be combined with the satellites to meet these special demands. Here in this section we present the major reasons for their conjugation. In Fig. 1 we show a satellite based IoT network. The coverage area of the satellite has been shown as an oval within which lies the IoT network. The IoT trans-receivers are the triangular towers and the red dots are the sensor nodes. Both the IoT and the satellite links are wireless.

### A. Motivation for Satellite IoT Integration

*1) Reliability in the wireless applications*
Reliability of wireless communications is still a challenging issue. Normal wireless networks do not provide the high reliability required in the mission critical applications. These issues can be addressed through the satellite networks. Satellites are normally available with a better reliability than the cellular networks. With proper constellation arrangements satellites provide more than 99.9% availability which is much higher than the current cellular networks. This is essential for the mission critical applications such as disaster management and military communications. High availability ensures high reliability under diverse conditions.

*2) Larger and broader coverage of the IoT Networks*
IoT networks are generally deployed over the existing cellular infrastructure. The cellular networks are oriented according to the human presence. Therefore the coverage of the cellular networks are very much limited. In case of the very high towers the coverage radius is around 20 km. High towers are not suitable for green applications as their radiation densities are high. However, satellite networks can cover a large area and they do not discriminate any part of the terrain. Thus, for a large scale multi-purpose deployment of IoT satellite based IoTs are preferred over the cellular IoTs.

*3) Better Security and Protection*
SIoT networks are very much oriented according to the presence and availability of the satellites. Intercepting and manipulating these systems are very rare and needs a lot of effort. In comparison to this, the cellular networks are very much compromised in terms of security and isolation. In addition to the above, satellite networks have added security measures such as anti-jamming mechanisms and strong cryptographic coding which cannot be manipulated easily.

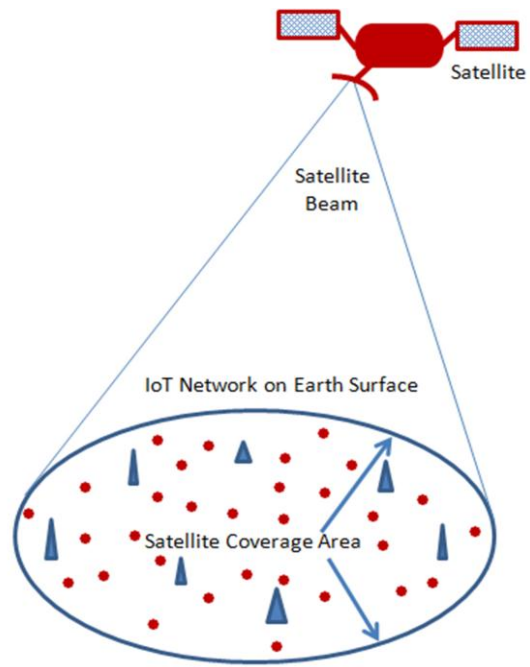

Fig. 1. Satellite based IoT Network.

*4) Economical and faster rural deployment:*
Economics plays an important role in every communication technology. In case of IoT, it has even more important roles as the numbers of ground terminals or the sensors are too many. Rural deployment of any IoT is very much dependent on the associated economy. SIoT is very promising in this regard as satellites have wider coverage and the satellite based IoTs need less resources than the cellular IoTs. Of course, the specific energy efficient versions of the IoTs would serve this sector better. We have addressed this aspect of SIoT later in section IV of this article.

*5) Multicasting of services:*
Each satellite has its own coverage area. These are dependent of their beam widths. Normally, the coverage area of a satellite is much larger than the cellular transmitters. Therefore multicasting through satellites is much easier and cost effective than the cellular transmissions. For the IoT based multicasting where the energy constraints are very much stringent it is very much beneficial.

*6) Longevity of the wireless networks*
In general the longevity of the satellite constellations is much higher than the cellular networks. The main problem with the cellular networks in this regard is the changing situations. Every decade several changes come into picture in the cellular networks. In the recent decades we see a new generation of mobile communication is replacing the existing ones. However, the satellites networks remain active for more than two to three decades. Therefore, the longevity of satellite networks ensures its supported ground networks too have a longer presence than the cellular networks.

*B. Satellite Cellular Dual Mode IoT*

There are several applications which need constant support from the backbone network. In such cases, the availability of the backbone is essential all the times. In case of the wireless cellular networks the availability is not very good. One of the major concerns in this regard is the dependence on the external power sources. Thus having a backup through some other means is essential for the critical applications. In such cases satellite support can be extended when cellular services are absent. These IoTs are known as satellite cellular dual mode IoT. Mission critical applications normally need the dual mode or a dedicated SIoT service.

III. APPLICATIONS OF SATELLITE BASED IOT SYSTEMS

There are several applications of the SIoT networks. Some of them are the existing ones and the rest are emerging in the changing scenarios. Right now the major focus is on the critical applications where the availability of the network has to be more than 99%. In addition to that other advantages of satellite networks such as larger coverage play the main roles in the SIoT applications.

*A. Mission Critical Applications*

Mission critical applications are the emergency and important services which need the availability of the networks all the time. Examples of mission critical applications are: disaster management during calamities, military data collection and transmission, real time control of important systems. These systems cannot afford the absence of the network for even a short time. During the natural disasters such as floods, cyclones, hurricanes and landslides, the wireless infrastructures are directly affected. Very often the networks go out of operation for days and in some cases even for months. However, the satellites are immune to these problems. That is why satellite based IoTs are preferred for the situation when the infrastructure is vulnerable to the natural disasters.

*B. Location Based Serivices*

Location based services need the exact locations to execute or deliver their services. IoT based networks along with the support of the satellites provide better location details than the cellular networks [10] – [16]. These fine details can be used for military applications and several other cases where location accuracies are critical. Global positioning systems using SIoT is preferable over other available alternatives due to their accuracies.

*C. Navigation Systems*

Navigation systems need the exact information of the surroundings both in the air and water. This information can be provided through SIoT. Coordination of the satellites with the support from the ground based IoT sensors provides much better accuracy than the satellites alone. In the new navigation systems SIOT can play a game changing role.

*D. Agricuture*

Agriculture is an important sector where the increase of yields is always a goal for the farmers. IoTs can be used for the precision farming. The information needed for agriculture can be collected through the SIoT networks. These information then can be utilized through the satellite backhaul to the SIoT sensors to monitor the water level, temperature and several other parameters.

*E. Tracking*

Tracking has several applications in many different sectors. For instance, in logistics tracking is used to provide the progress in the transportation or delivery. Planes ae tracked for their navigation and safety. Animal tracking is used to know the location of the cattle in the fields, their health monitoring and to avoid their theft. Similarly, kids are tracked by the parents for their safety and security. Police tracks the criminals and thieves for the tasks related to law enforcement. These tracking mechanisms need both the satellites and IoT networks for better accuracies. SIoT can play vital roles in tracking.

*F. Healthcare*

Healthcare sector is fundamental for every country. It is a basic need for every citizen. There are several aspects of healthcare which needs the support of SIoT. For instance remote health monitoring and premedical treatment of the patients before they arrive in the hospital can be provided through SIoT networks. This is even more critical where there are no cellular networks.

There are many applications of SIoT, all of them we cannot present here. Many new applications emerge every year. Based on these demands for SIoT many companies have been established who provide services in any different forms using SIoT. LEO satellites are popular for the SIoT applications. That is because they are nearer to the earth surface. The power budget for the LEO satellites is lower than the geostationary satellites.

IV. SUSTAINABILITY OF SATELLITE BASED IOT NETWORKS

Satellite IoT combination is a sophisticated network. Satellites these days are used for multiple purposes. Therefore, SIoT applications through the satellites will be just another set of applications for the satellites. The sustainability of the SIoT has two segments. One is associated with the ground, mainly the IoT segment and the other is associated with the space around the satellite. Overall SIoT networks need the proper coordination between both the segments for their long term sustainability. Energy wise there are different types of IoTs which need the transmitters and receivers of different specifications. These days energy efficient IoTs are available and even very low energy consuming IoTs are becoming popular. However, for proper satellite linking, the received amounts should be more than the threshold levels of the IoT receivers.

*A. Power budget*

Power required for the IoTs to remain connected with the satellites should have the following considerations. The sensors of the IoTs need a minimum power for their operation which is known as the threshold power. There will be losses in the channel from the satellites to the sensors. In addition to these aspects the noise margin and system margins should be considered. Based on these considerations we have presented equation (1) for the power budget.

$$P_T = P_L + P_{IoT,Th} + P_{NM} + P_{SM} \qquad (1)$$

Here, $P_T$, $P_L$, $P_{IoT,Th}$, $P_{NM}$ and $P_{SM}$ are the power transmitted by the satellite, power loss along the channel, threshold power needed by the IoT device, noise margin and system margin respectively. It is noteworthy that the satellites used for SIoT are mainly the LEO satellites which are not stationary over a specific location of earth and their distances keep changing with time. Therefore the power parameters shown in (1) are not constant, rather they too keep on changing depending on the locations of the satellites from the IoT networks. Except for $P_{IoT,Th}$, in the right hand side the rest three parameters are variables. Therefore, $P_T$ can be varied according to the parameters of the right hand side or it can be maintained at a high enough value so that $P_{IoT,Th}$ is always available at the IoT sensor terminals.

*B. Narrowband SIoT Networks*

Narrowband versions of IoTs are available and their standards too have been ready. NBIoT, was first standardized in LTE Release 13. According to this, it has two transmission levels: 20 dBm and 23 dBm [6]. The receiver power level can be as low as -64 dBm. NBIoTs take a very small amount of power and bandwidth for their operations. NBIoT sensors can be powered through small batteries for years. Even energy harvesting mechanisms can be coupled with its sensors for green operations. This is a low power wide area network (LPWAN). Therefore very much suitable for the rural and energy limited applications [6]. Combination of NBIoT with satellite networks can be very much attractive for several energy limited applications. Furthermore, NBIoT can also provide enhanced services in specific cases. In LTE Release 14, these new specifications for enhanced services have been mentioned.

Overall, SIoT can be made sustainable through the green technologies. NBIoT and several other LPWAN are suitable for the conjugation with satellites for sustainable SIoT. This is really good for rural areas in the developing countries where cost of deployment plays the main role.

V. CONCLUSIONS

In this paper, we presented the main utilities and motivation for SIoT. There are several new applications emerging for SIoT. Majority of these applications normally do not have a suitable alternative. Therefore, SIoT are critical in several emerging applications. In order to make SIoT sustainable and efficient there are some specific forms of IoT such as the NBIoT. It can make SIoT green. The commercial perspectives of SIoT are quite strong. Several large companies have integrated SIoT in their businesses. In addition to that many startups too have been established for SIoT based services.

REFERENCES

[1] L. Da Xu, W. He, and S. Li, "Internet of Things in industries: A Survey," IEEE *Transactions on Industrial Informatics*, vol. 10, no. 4, pp. 2233 – 2243, Nov. 2014.

[2] A. Al-Fuqaha, M. Guizani, M. Mohammadi, M. Aledhari, and M. Ayyash, "Internet of Things: A survey on enabling technologies, protocols and applications," *IEEE Communication Surveys Tutorials*, vol. 17, no. 4, pp. 2347 – 2376, Nov. 2015.

[3] S. Mohanty, and S. K. Routray, "CE-Driven Trends in Global Communications: Strategic Sectors for Growth and Development," *IEEE Consumer Electronics Magazine*, vol. 6, no. 1, pp. 61 – 65, Jan. 2017.

[4] D.Hu, L. He, and J. Wu, " A Novel Forward-link Multiplexed Scheme in Satellite-based Internet of Things," *IEEE Internet of Things Journal*, vol. 5, no. 2, pp. 1265 – 1274, Feb. 2018.

[5] Z. Qu, G. Zhang, H. Cao, and J. Xie, "LEO Satellite Constellation for Internet of Things," *IEEE Access*, vol. 5, pp. 18391 – 18401, 2017.

[6] S. Routray, K. P. Sharmila, "Green Initiatives in IoT" in Proceedings of IEEE International Conference on Advances in Electrical, Electronics, Information, Communication and Bio-Informatics (AEEICB), Chennai, India, 27-28 Feb. 2017.

[7] C. Zhu, V. C. M. Leung, L. Shu, and E. C.-H. Ngai, "Green Internet of Things for Smart World," *IEEE Access*, vol. 3, no. 11, pp. 2151-2162, Nov. 2015.

[8] Overview of NBIoT. [Online] (retrived on 15 June 2018). Available: https://www.link-labs.com/blog/overview-of-narrowband-iot.

[9] J. Guo, H. Zhang, Y. Sun, and R. Bie, "Square-root unscented Kalman filtering-based localization and tracking in the internet of things," *Personal and ubiquitous computing*, vol. *18, no.* 4, pp.987 – 996, Apr. 2014.

[10] Y. Gu, and F. Ren, "Energy-Efficient Indoor Localization of Smart Hand-Held Devices Using Bluetooth," *IEEE Access,* vol. 3, no. 6, pp. 1450 – 1461, Jun. 2015.

[11] A. Colombo, D. Fontanelli, D. Macii, and L. Palopoli, "Flexible indoor localization and tracking based on a wearable platform and sensor data fusion," *IEEE Transactions on Instrumentation and Measurement*, vol. *63, no.* 4, pp.864-876, Apr. 2014.

[12] Z. Chen, F. Xia, T. Huang, F. Bu, and H. Wang, "A localization method for the Internet of Things," *The Journal of Supercomputing*, pp.1-18, 2013.

[13] C. Wang, M. Daneshmand, M. Dohler, X. Mao, R. Q. Hu, H. Wang, "Guest Editorial-Special issue on internet of things (IoT): Architecture, protocols and services," *IEEE Sensors Journal*, vol. *13, no.* 10, pp. 3505 – 3510, Oct. 2013.

[14] S. Ramnath, A. Javali, B. Narang, P. Mishra, S, K, Routray, "IoT Based Localization and Tracking,"in Proc. of IEEE International Conference on IoT and its Applications, Nagapattinam, India, May, 2017.

[15] S. Ramnath, A. Javali, B. Narang, P. Mishra, S, K, Routray, "An Update of Location Based Services,"in Proc. of IEEE International Conference on IoT and its Applications, Nagapattinam, India, May, 2017.

[16] D. Minoli, "M2M Developments and Satellite Applications," in Innovations in *Satellite Communications and Satellite Technology, The Industry Implications of DVB-S2X, High Throughput Satellites,* pp. 221 – 296, 2015.

[17] S. K. Routray, M. K. Jha, L. Sharma, R. Nymangoudar, A. Javali, and S. Sarkar, "Quantum Cryptography for IoT: A Perspective," in Proc. of IEEE International Conference on IoT and its Applications, Nagapattinam, India, May, 2017.

[18] Advantages of SIoT. [Online] (retrived on 15 June 2018). Available: https://blog.orbcomm.com/why-satellite-for-iot-6-key-advantages/.